\documentclass[12pt]{article}

\usepackage{latexsym,amsmath,amssymb,theorem,epsfig}

\usepackage{graphicx}
\usepackage{graphicx,subfigure}
\usepackage{epstopdf}
\usepackage[body={17.5cm, 21cm},right=2cm]{geometry}
\usepackage{amssymb}
\usepackage{amsmath}
\usepackage{psfrag}
\usepackage{epsfig}
\usepackage{cancel}
 \allowdisplaybreaks[4]

\usepackage[all]{xy}

\DeclareFontFamily{OT1}{rsfs10}{}
\DeclareFontShape{OT1}{rsfs10}{m}{n}{ <-> rsfs10 }{}
\DeclareMathAlphabet{\mathscript}{OT1}{rsfs10}{m}{n}

\numberwithin{equation}{section}

\newcommand{\ns}{\normalsize}

\newcommand{\cH}{{\cal H}}

\newcommand{\be}{\beta}

\def\gsim{ \lower .75ex \hbox{$\sim$} \llap{\raise .27ex \hbox{$>$}} }
\def\lsim{ \lower .75ex \hbox{$\sim$} \llap{\raise .27ex \hbox{$<$}} }
\def\be{\begin{equation}}
\def\ee{\end{equation}}
\def\bea{\begin{eqnarray}}
\def\eea{\end{eqnarray}}

\begin{document}

\begin{titlepage}

\vspace{-5cm}

\title{
  \hfill{\ns }  \\[1em]
   {\LARGE   The IR-Completion of Gravity:  \\[1mm] What happens at Hubble Scales?}
 }

\author{
   Federico Piazza\footnote{fpiazza@perimeterinstitute.ca}
     \\[0.5em]
   {\ns Perimeter Institute for Theoretical Physics}\\
{\ns Waterloo, Ontario, N2L 2Y5, Canada}}

\date{}

\maketitle

\begin{abstract}

We have recently proposed an ``Ultra-Strong" version of the Equivalence Principle (EP) that is not satisfied by standard semiclassical gravity. In the theory that we are conjecturing, the vacuum expectation value of the (bare) energy momentum tensor  is exactly the same as in flat space: quartically divergent with the cut-off and with no spacetime dependent (subleading) terms.  The presence of such terms seems in fact related to some known difficulties, such as the black hole information loss and the cosmological constant problem. Since the terms that we want to get rid of are subleading in the high-momentum expansion, we attempt to explore the conjectured theory by ``IR-completing"  GR.
We consider a scalar field in a flat FRW Universe and isolate the first IR-correction to its Fourier modes operators that kills the quadratic (next to leading) time dependent divergence of the stress energy tensor VEV. 
Analogously to other modifications of field operators that have been proposed in the literature (typically in the UV), the present approach seems to suggest a breakdown (here, in the IR, at large distances) of the metric manifold description. We show that corrections to GR are in fact very tiny, become effective at distances comparable to the inverse curvature and do not contain any adjustable parameter. Finally, we derive some cosmological implications.
By studying the consistency of the canonical commutation relations, we infer a correction to the distance between two comoving observers, which grows as the scale factor only when small compared to the Hubble length, but gets relevant corrections otherwise. The corrections to cosmological distance measures are also calculable and, for a spatially flat matter dominated Universe, go in the direction of an effective positive acceleration.

\end{abstract}

\thispagestyle{empty}

\end{titlepage}

\section{Introduction}

Modifications of General Relativity (GR) on the largest scales have been advocated in order to give account for the present acceleration of the Universe. Attempts in this direction \cite{massive} typically maintain the metric manifold structure of GR  and change the dynamics of the gravitational field by effectively giving the graviton a small mass. It is thoroughly assumed that any infra-red (IR) modification of gravity must bring in at least one new mass parameter,  corresponding to the scale at which such a modification becomes effective. However, that is not necessarily the case. 
In \cite{ultra} we put forward an infra-red modification of GR that does not involve any new adjustable parameter but that might require a substantial reassessment of the present understanding of gravity in the semi-classical regime. In this paper we review and refine the model sketched in \cite{ultra} and derive some of its cosmological implications.

We start from usual semi-classical gravity, where matter fields are quantized on a curved background manifold, and we modify in the IR the operators of the matter field theory.  In a cosmological setup, operators corresponding to Fourier modes of physical momentum $k$ are corrected by order $H^2/k^2$, $H$ being the Hubble parameter. The coefficient of such a modification is fixed by imposing an ``Ultra-Strong" version of the equivalence principle, which we are going to recall in the following. We argue that 
such IR-modified operators fail to be supported by a metric manifold structure at large scales. Hence,  the proposed modification seems to indicate a breakdown, for systems of size comparable to the inverse curvature, of the geometrical description of space-time as a metric manifold. 

It is known that there is a strict connection between the geometric properties of a manifold and the spectrum of the differential operators \cite{heat} or the algebra of functions \cite{connes} therein defined; such abstract characterizations have occasionally been used for generalizing common geometrical concepts and the description of spacetime itself~\cite{connes,sw}. However, so far, attempts in this direction have always been applied to the UV and intended to modify spacetime at the smallest scales. 
Although perhaps less intuitive, a modification effective in the IR -- and at length scales set by the curvature -- is appealing on both theoretical and observational grounds:
\begin{itemize}

\item The standard paradigm 
``GR + matter fields" presents  some difficulties that are arguably UV-insensitive,
such as the cosmological constant problem \cite{wein} and the black hole information paradox \cite{haw}. Such difficulties appear already at the semi-classical/low-energy effective level and therefore it seems unlikely that any modification at the Plank scale or UV-completion of the theory might possibly address them. Questioning the IR-side of the picture seems thus a valuable alternative.

\item The apparent acceleration of the Universe is described in GR as something kicking in at some late-time epoch in the cosmological evolution and related to an extremely tiny mass scale. Such a scale is adjusted \emph{ad hoc} in the models of ``Dark Energy", which, therefore, generically suffer from fine-tuning and ``coincidence" problems. 
The alternative view suggested here is based on the simple remark that any high redshift object is placed from us at distances of order the inverse curvature (i.e. the Hubble radius $H_0^{-1}$). Hence,  modifying the usual framework at a length scale set by the curvature -- rather than fixed \emph{a priori} --  will systematically affect any cosmological observation at high redshift, regardless of when such an observation takes place and without the need of any external mass parameter.

\end{itemize}

Before summarizing the cosmological implications of the proposed deformation let us review the main theoretical idea: the Ultra Strong Equivalence Principle~\cite{ultra}. Consider the vacuum expectation value of the local energy density of a (massless) field according to standard semi-classical gravity; this can be expanded at  high momenta as 
\begin{eqnarray} \label{structure}
\langle T_0^0(x,t)\rangle_{\rm bare} &=& \int d^3 k\left(k  + \frac{f_{\rm quad}(t)}{k} + \frac{f_{\rm log}(t)}{k^3} + \dots  \right) \\[2mm]
& = &\ \  {\rm local\ terms}\ \ +\ \ {\rm non \ local\ terms}. \nonumber
\end{eqnarray}
Here spatial homogeneity has been assumed for simplicity and the $f$s are functions of time of appropriate dimensions (see eq. \eqref{square} below, or reference \cite{fulling2}, for the explicit expression in the case of a massive scalar in a flat FRW Universe). The stress tensor renormalization is a well-defined prescription to make sense of the infinities appearing in \eqref{structure}: the local terms can be subtracted by local gravitational counterterms (cosmological constant, Newton constant and higher order local operators). The finite non-local pieces left represent the genuine particle/energy content of the choosen ``vacuum" state. 

It is interesting to note that some of the difficulties afflicting the standard low-energy framework for gravity seem to be related to the spacetime dependent subleading terms (the $f$s in the case considered above) in the energy momentum VEV expansion. The black hole information loss paradox is related to the presence of the non-local pieces, as they are the ones responsible for particle creation and for the evaporation of the black hole. We argue that also the cosmological constant problem is related to the presence of those terms, at least if we restrict the analysis to the level of non-quantized semi-classical gravity. Such a problem appears in the renormalization procedure that we described earlier as the huge amount of fine tuning that is required to match the first quartically divergent piece of \eqref{structure} with its observed value. However, the quartically divergent term is not particularly worrisome by itself, essentially because it is a constant. In flat space we deal with it, by normal ordering or just by subtracting a constant, albeit infinite, contribution. In other words, as brutal as it might appear, ``the vacuum does not gravitate" is a well defined prescription, as long as the contribution to be neglected is a spacetime-independent scalar. We are arguing that 
the main source of trouble is not the first term in \eqref{structure}, but  the remaining spacetime dependent pieces, as they make the normal ordering prescription meaningless and the renormalization procedure necessary in the first place.

Based on the above reasoning and encouraged by few (more philosophical) arguments explained below in Sec.~\ref{sec2}, the  conjecture at the basis of this paper is that 
\emph{there exists a theory that resembles standard semiclassical gravity at small scales but in which, as in flat space, all time-dependent pieces of \eqref{structure} (the $f$s) at any point-event, and for some ``vacuum" state, just do not exist}. 
Since the terms that we want to get rid of are effectively IR with respect to the first quartically divergent part and can be calculated, we should be able to probe this theory by IR-completing standard semi-classical gravity. 

It is clearly not guaranteed that a consistent theory with the required properties exists in full generality beyond the example considered here; if it does, its formulation will certainly require a major theoretical breakthrough. Giving up the metric manifold structure -- and dealing only with quantum operators -- constitutes a significant technical difficulty. On the other hand, compared to some of the subtleties of standard QFT on curved spacetime~\cite{birrell}, the physical picture that we are insinuating is somewhat simplified, made trivial. We are conjecturing the possibility of describing, locally, the physical phenomena in exactly the same way as we do in flat space, and that  the large scale-properties of spacetime have no effect whatsoever inside Einstein's free-falling elevator.
This is the spirit of the Equivalence Principle, that we are upgrading to the level of semi-classical gravity and making ``Ultra-Strong"~\cite{ultra}. The local resemblance to flat space is pushed to the level of the quantized fields, whose energy momentum tensor VEV at any point-event is conjectured to amount, for some unambiguously chosen vacuum state, to the constant value  $\sim \int k d^3 k $.

In this paper we try to explore the conjectured theory through a sort of Taylor expansion around GR, 
the zeroth order being a flat FRW Universe with a given scale factor $a(t)$ and a spectator free scalar field in it. We isolate the first correction to the scalar field Fourier operators that reabsorbs the quadratic divergence $f_{\rm quad}$ in \eqref{structure}. In a similar fashion, if we did not know GR, we might at least try to explore it 
by expanding around flat space at a given point in Riemann normal coordinates. 
The proposed modification is first introduced, in Sec. \ref{sec4.5}, through modified commutation relations between the global creators and annihilators of the field,
\begin{equation}
[A_{\vec{n}}^{(1)}, A_{\vec{n}'}^{(1)\, \dagger}] = \delta^3(\vec{n} - \vec{n}')\left(1 - \frac{H^2 a^2}{2 n^2} + {\rm higher\ orders}\, \right).
\end{equation}
In the above, $\vec{n}$ is the comoving momentum and 
the form of the correction term is dictated by the Ultra-Strong Equivalence principle.
At this order of approximation, however, there seems to be another more practical way to see the proposed modification, which we show in Sec. \ref{sec5}.
After a simple rescaling, one can deal with the usual ``manifold-operators", labeled by the ``manifold-momenta" $\vec{n}$, that satisfy the usual commutation relations, $[A_{\vec{n}}, A^\dagger_{\vec{n}'}] = 
\delta_{{\vec{n}},{\vec{n}}'}$. The proposed modification then boils down to introducing a slight mismatch between the ``global, manifold" label $\vec{n}$ and the physical (comoving) momentum $\vec{k}$ that locally defines the infinitesimal translations and the derivatives of the local fields:
\begin{equation}\label{kintro}
\vec{k} = \vec{n} \left(1 - \frac{H^2 a^2}{2 n^2}\right).
\end{equation}
Note that the comoving physical momentum is time dependent in this model, although only negligibly at sub-Hubble scales.

The modified dispersion relations \eqref{kintro} come from requiring the cancellation of the quadratically divergent time-dependent piece $f_{\rm quad}(t)$ in the high momentum expansion \eqref{structure} at some given point ``$\vec{x}=0$". In Sec. \ref{secglobal} we try to make sense, globally, of such modified dispersion relations and explore this spacetime a finite distance away from $\vec{x} = 0$ by using the accordingly modified translation operator. As much as the physical comoving momentum \eqref{kintro} is conserved only approximately, 
 it turns out from the analysis of Sec. \ref{secglobal} that the physical distance $d$ between two comoving observers grows as the scale factor $a(t)$ only if that distance is much smaller than Hubble.  At large scales, the expansion rate effectively detaches from the local expansion $a(t)$. This is the most striking signal of a breakdown of the manifold description. The comoving distance $\lambda = d/a(t)$ of two faraway comoving observers, instead of being constant, satisfies
\begin{equation}\label{exintro}
\dot{\lambda} = \lambda^3 \frac{(H^2 a^2)\dot{}}{4} + {\rm higher\ orders}.
\end{equation}
The effect is clearly negligible  at sub-Hubble scales but will affect the observed cosmological dynamics at high redshift. 

The cosmological implications/predictions of the present model, obtained by applying the proposed IR modification to a spatially flat FRW Universe, are derived and discussed in Sec. \ref{seccosm} and can be summarized as follows:
\begin{itemize}
\item 
Comoving momenta $\vec{k}$ are conserved only when they are very high compared to Hubble, i.e. when $k \gg a H$. More generally, what is conserved is the quantity

\begin{equation} \label{conserved}
\vec{k} \left(1 + \frac{H^2 a^2}{2 k^2} + {\rm higher\ orders}\right).
\end{equation}

\item The physical distance $d$ between two comoving observers grows in time proportionally to the scale factor $a(t)$ only if the two are well within the Hubble radius. More generally, their distances  at two different times $t'$ and $t$ are related by 

\begin{equation} \label{exintros}
\frac{d(t)}{d(t')} = \frac{a(t)}{a(t')}\left[1 +  \frac{d^2(t')}{4}\left(H^2(t) \frac{a^2(t)}{a^2(t')} - H^2(t')\right) + {\rm higher\ orders} \right].
\end{equation}

\item
The angular-diameter distance is given, in terms of the redshift, by

\begin{equation} \label{lumiintro}
d_A(z) = \frac{r(z)}{1+z}.
\end{equation}
In a matter dominated Universe, $r(z)$ is the solution of
\begin{equation} \label{rintro}
\frac{d (H_0 r)}{dz} = \frac{1}{(1+z)^{3/2}} +  \frac{(H_0 r)^3}{4} + {\rm higher\ orders}.
\end{equation}
with initial condition $r(0) = 0$.
\item
The luminosity distance $d_L$, at least at this order of approximation, is related to $d_A$ in the usual way, $d_L = (1+z)^2 d_A$.
\end{itemize}
The cosmological distances derived with equation \eqref{rintro} in a matter dominated Einstein-de Sitter Universe are corrected in the direction of an apparent ``acceleration" and show some resemblance to those of a $\Lambda$CDM Universe  (Figs. \ref{fig1} and \ref{fig2}, Sec. \ref{seccosm}). A more quantitative analysis will follow \cite{savvas}. 

The paper is organized as follows. The general motivations and the main idea are given in Secs. \ref{sec2} and \ref{sec3} which are basically taken from the shorter note \cite{ultra}. In Sec. \ref{sec4} we review the standard calculation of \eqref{structure} for a free scalar in a flat FRW Universe. In Sec. \ref{sec4.5} we modify the Fourier operators in order to get rid of the quadratic divergences there encountered. Modified dispersion relations are introduced in Sec. \ref{sec5} as an essentially equivalent, but more practical, way to look at the proposed modification. In Sec. \ref{secglobal} we attempt to make sense, globally, of the proposed modification and study its implications on large scales. In Sec. \ref{seccosm} we finally derive some cosmological consequences, such as the modified luminosity distance. Potential problems and future perspective are finally summarized in Sec. \ref{sec9}.

\subsection{The Equivalence Principle: Gravity as an IR effect.}\label{sec2}

The equivalence principle (EP) can be formulated simply as follows: \emph{inside a sufficiently small free-falling elevator you do not see the (classical) effects of gravity}. Amusingly, such a cornerstone of modern physics is actually stating what  (where) gravity is not, rather than what  (where) gravity is!  Among the many celebrated implications of general relativistic physics, the view that we aim to stress here is that EP forces us to consider and describe gravity as an IR phenomenon, whose effects are visible only outside the free-falling elevator. How EP turned into a consistent theory is well known: gravity is beautifully encoded in GR as the geometry of the physical space-time and therefore its effects are automatically suppressed within those systems that are much smaller than the inverse curvature. 
By changing (curving) the large-scale structure of spacetime, GR leaves the smallest systems free of classical gravitational effects. Notably, the  IR scale where non-gravitational physics breaks down is not a parameter of the theory, but is set by the local curvature $R$. Schematically, in three dimensions, the area of a two-sphere of radius $l$ and volume $V$ receives corrections from the flat-space expectation of the type
\begin{equation} \label{1}
A(l) \ = \ 4 \pi l^2 \, (1 + {\cal O}(l^2 R)) \ =\  (36 \pi V^2)^{1/3}\, (1 + {\cal O}(R V^{2/3}))  .
\end{equation}

The effects of gravity, originally banned from the free-falling elevator, reappeared, after the developments of quantum theory, through what one might call the back door.
The fields quantized on a curved manifold are sensitive to the global properties of spacetime because their modes are defined on the whole of it. As a result, inside the elevator, you will generically experience, and possibly detect with your local instruments\footnote{It is arguable that a small system cannot detect quanta of arbitrarily small energies. While this is generically true in practice, there is, to our knowledge, no universal (independent of the details of the matter field dynamics) and in principle argument that limits the possibility of particle detection by small detectors. (On the opposite, there is a very precise and quantitative sense in which
tidal forces in GR become small within small systems.) Within a system of size $L$, the first excited level is typically at energy $1/L$ above the ground state and cannot be excited by photons of much lower energies. However, the structure of the higher energy meta-stable states is often very rich, and energy differences among such states can be arbitrarily small. An atom, for example, has extremely fine energy transitions that, in principle, allow to ``detect" photons of wavelengths much larger than its size.}, particle creation because of non-local gravitational effects. 
Clearly, the process of quantum particle creation does not contradict EP, which was formulated within the framework of classical physics. Nevertheless, it is tempting to try a different path than the one historically followed: rather than imposing field quantization on top of a curved manifold, here we attempt to upgrade the equivalence principle and extend it to the quantum phenomena.
Thus, we will consider a stronger version of EP, in which all the effects of gravity are definitely forbidden inside the elevator, including the quantum effects that in the standard semi-classical treatment lead to particle creation. Effectively, what we want is that 
\emph{for each matter field or sector sufficiently decoupled from all other matter fields, there exists a state, the ``vacuum", that is experienced as empty of particles by  each free-falling observer.}

Following the discussion given in the introduction, a possible concrete strategy would be 
trying to reabsorb in the IR the non-local terms inside \eqref{structure} after the stress energy tensor has been renormalized by usual means. However, we will follow a different, more radical, plan. We will circumvent the usual procedure of stress-tensor renormalization and just try to cancel the time-dependent terms  of the \emph{bare} energy momentum tensor (all terms inside \eqref{structure} except the first). In fact, the time-dependent pieces can arguably be reabsorbed with some IR modification, because they are effectively IR with respect to the quartically divergent term.
Moreover, bypassing the usual procedure of stress tensor renormalization might, at the same time, shed some new light on the cosmological constant problem, at least at the level of non-quantized semi-classical gravity. If, in the IR-completed theory,
all the time dependent terms just do not exist, then we do not need to renormalize the stress tensor anymore. Of course, we are still left with the leading spacetime-constant quartic divergence, but we can live with it and treat it, as we do in flat space, by normal ordering.   Finally, it is interesting to reabsorb the  quadratic divergence in the IR, rather than with a local counterterm, because the required modification, as we are going to show, is of the right order of magnitude to give interesting cosmological implications. 

\begin{quote}
{\bf Ultra Strong Equivalence Principle:}
For each matter field or sector sufficiently decoupled from all other matter fields, there exists a state, the ``vacuum", for which the expectation value of the (bare) energy momentum tensor reads the same as in flat space, regardless of the configuration of the gravitational field.
\end{quote}

As a -- rather obvious -- warning to the reader we mention that, while EP is extremely well tested (see e.g. \cite{torsion}), the proposed ultra strong version is not. Experimental hints might actually be pointing against it: the mechanism of quantum particle creation
is successfully exploited by inflationary models in order to produce the primordial spectrum of  cosmological fluctuations.
The present knowledge of the earliest Universe is, however, too model dependent (and largely based on the assumption that such a mechanism is at work) for that to be considered as an experimental evidence.

\section{Regions of Space as Quantum Subsystems} \label{sec3}

Before describing the model, it is worth examining closely what happens in the standard formulation of semi-classical gravity, where fields are quantized on a curved background manifold. Consider a spacetime with a global time foliation labeled by a time parameter $t$. The Universe as a whole at time $t$ is a three-dimensional manifold ${\cal M}$ in the GR description. The matter quantum fields are instead described by a quantum state living in a Hilbert space $\cH$. Thus, if now we consider a region of space (at time $t$) of finite volume $V$ (``this room, now"), that has two complementary descriptions \cite{fedo1,fabio1,sergio}: it is a \emph{submanifold} according to GR and a \emph{quantum subsystem} for the quantized fields. 
The correspondence sub-manifold/sub-system is explicitly realized by  the set of local operators $A(t, x)$ of the field theory. Those act on the quantum system $\cH$ but have labels $x$ living on the three-dimensional manifold ${\cal M}$. As a consequence, we can take integrals of scalar local operators over some region of volume $V$,
\begin{equation} \label{bah}
A(t, V) = \int_V d^3 x \sqrt{-g^{(3)}}\ A(t, x) ,
\end{equation}
which are still operators acting on the total Hilbert space $\cH$. More precisely, $A(t,V)$  acts non-trivially only on the quantum subsystem $\cH_V$ corresponding to the region of space that has been integrated over, and as the identity on the rest of the system $\cH_{\rm Rest}$. In this way, the algebra of operators $A(t, V)$ determines \cite{paolo} the partition $\cH = \cH_V \otimes \cH_{\rm Rest}$ of the quantum system/Universe and therefore can be taken as a definition of the region of space $V$. Defining regions of space as quantum subsystems \cite{fedo1} looks like a useless complication in the standard formalism because the correspondence  sub-system/sub-manifold is always implicitly at work: such a correspondence is set once and for all by \eqref{bah}, given the set of local operators $A(t,x)$.

When we say that the metric-manifold description might break-down, we mean, more precisely, that regions of space, still perfectly defined as quantum subsystems, may not have the nice property that the corresponding operators integrate as in \eqref{bah}.
We argue therefore that \eqref{bah} is valid only in the limit of zero curvature and that operators corresponding to extended regions of space can be written as \eqref{bah} only up to order ${\cal O}(R V^{2/3})$, where $R$ is some curvature scalar. Here is the mnemonic thumb-rule that will guide us in our further developments:  
\begin{equation} \label{2}
A(t, V) \ \simeq \  \int_V d^3 x \sqrt{-g^{(3)}}\ A(t, x) \ \left[1 + {\cal O}(R V^{2/3})\right] .
\end{equation}
We are deliberately mimicking the general type of corrections \eqref{1} that non-extensive geometrical quantities undergo in the transition from flat space to curved space. The idea is to extend this type of behavior also to extensive quantities, that in the standard description are just proportional to the volume. The implications of \eqref{2} are quite striking. Consider, for instance, a
perfectly homogeneous Universe. By definition, each comoving observer  measures in its surrounding the same energy density $\rho$. Eq. \eqref{2} implies that in that Universe, if one starts considering regions of space of Hubble size, the total energy 
inside that region will drastically differ from the three-dimensional integral of the local densities measured by the observers living therein.
In a metric manifold, that would be a dramatically non-local effect.

\section{The Zeroth Order: a Flat FRW Universe} \label{sec4}

In this section we briefly review the quantization of a free scalar field in a FRW spatially flat  metric manifold. In the spirit of this paper, what we are considering here is therefore the zeroth order approximation of the theory that we are trying to explore. From the action
\begin{equation} \label{action}
S = \frac{1}{2}\int d^4 x \sqrt{-g}\left(\partial \phi^2 - m^2 \phi^2\right)
\end{equation}
and the metric element 
\begin{equation}
d s^2 = dt^2 - a^2(t)(d \vec{x}^2) = a^2(\tau) (d\tau^2 - d \vec{x}^2)
\end{equation}
we derive the equation of motion for the field, 
\begin{equation} \label{equation}
\ddot \phi(t, \vec{x}) + 3 H \dot \phi (t, \vec{x}) - \frac{\partial_i^2}{a(t)^2}  \phi (t, \vec{x}) + m^2 \phi (t, \vec{x}) = 0,
\end{equation}
 and its energy momentum tensor, $T^{\mu \nu} = \frac{2}{\sqrt{-g}}\frac{\delta S}{\delta g^{\mu \nu}}$, from which we get the expression of the energy density of $\phi$ at time $t$ and comoving spatial coordinate coordinate $\vec{x}$:
\begin{equation} \label{calH}
T^0_0(t, \vec{x}) = {\cal H}(t, \vec{x}) = \frac{1}{2}\left[{\dot \phi}^2 + \frac{1}{a^2} (\partial_j \phi)^2 + m^2 \phi^2\right].
\end{equation}
A dot means derivation with respect to the proper time $t$ while a prime denotes derivation 
with respect to conformal time $\tau$; $H = {\dot a}/a$ is the Hubble parameter. Upon quantization in the Heisenberg picture, the field can be expanded in creators and annihilators:
\begin{equation} \label{field}
\phi(t, \vec{x}) = \frac{1}{(2 \pi)^{3/2}}\int d n^3\left[\psi_n(t) e^{i \vec{n} \cdot \vec{x}} A^{(0)}_{\vec{n}} + \psi^*_n(t) e^{-i \vec{n} \cdot \vec{x}} A^{(0) \dagger}_{\vec{n}}\right].
\end{equation}
For reasons that will be clear in the following, we have called $\vec{n}$ (and not $\vec{k}$) the comoving momentum label in this FRW space; $\vec{n}$ is a conserved quantity, related to the proper physical momentum $\vec{p}$ by $\vec{p} = \vec{n}/a(t)$. The suffix
$(0)$ stands for ``zeroth order" since the creation operators just introduced are intended to be the zeroth order (--metric manifold) approximation of the operators of the IR-completed theory. 

From the action \eqref{action} one derives the conjugate momentum 
\begin{equation}
\pi(t, \vec{x}) = \frac{\delta S}{\delta \phi (t, \vec{x})},
\end{equation}
and from the canonical commutation relations 
\begin{equation} 
[\phi (\vec{x}), \pi(\vec{x}\, ')] = i\delta^3(\vec{x}-\vec{x}\, '),
\end{equation}
one derives the canonical commutators for the zeroth order-- global operators
\begin{equation} \label{normalcom}
[A^{(0)}_{\vec{n}}, A^{(0) \dagger}_{\vec{n} '}] = \delta^3(\vec{n} - \vec{n}\, ')\, .
\end{equation}
It is customary to choose $A^{(0)}_{\vec{n}}$ as the operator that always annihilate the vacuum. The chosen vacuum state is therefore implicitely characterized  by the choice of the 
mode functions $\psi_n(t)$, that, by \eqref{field} and \eqref{equation}, satisfy
\begin{equation} \label{modes0}
\ddot \psi_n + 3 H \dot \psi_n +  \omega_n^2 \psi_n = 0 ,
\end{equation}
where
\begin{equation}
\omega_n = \sqrt{\frac{n^2}{a^2} + m^2} = \sqrt{p^2 + m^2}
\end{equation}

There is an extensive literature about the correct vacuum choice, i.e., the appropriate solutions of \eqref{modes0}. The classical argument in favor of the \emph{adiabatic vacuum} \cite{birrell,fulling,fulling2} is its resemblance to the flat space behavior for those modes that are deep inside the Hubble scale. Within the standard framework, the adiabatic vacuum is as close as we can get to implement the general ideas described in the previous sections and, therefore, will be our starting point. The adiabatic vacuum corresponds to the choice
\begin{equation} \label{adiabatic}
\psi_n(t) = \frac{1}{\sqrt{2 a^3 W_n}} \exp \left(-i \int^t W_n dt'\right),
\end{equation}
where the function $W_n(t)$ can be approximated at arbitrary adiabatic order with a formal WKB expansion. For the next to leading order in the momentum $n$ expansion we need no more than the next to leading adiabatic order \cite{fulling,fulling2}:
\begin{equation} \label{adiabatic2}
W_n(t) = \omega_n(t)(1+ \epsilon_2)^{1/2},
\end{equation}
where the function $\epsilon_2$ can be derived from a WKB expansion and amounts to 
\begin{equation} \label{eps2}
\epsilon_2 = -\frac{a^3}{(a^3 \omega_n)^{3/2}}\frac{d}{d t}\left[\frac{1}{\omega_n}\frac{d}{d t}(a^3 \omega_n)^{1/2}\right] = -\frac{a^2}{n^2}\left(\frac{\ddot a}{a} + H^2\right)\left[1 + {\cal O}\left(\frac{m^2}{n^2}\right)\right].
\end{equation}

The vacuum expectation value of the energy density at a given point is calculated straightforwardly:
\begin{eqnarray} \label{bahbah}
\langle 0 |\, T_0^0 (t, \vec{x}) \, | 0 \rangle & = & \frac{1}{2(2\pi)^3}
\int d^3 n \, d^3 n' \left[{\dot \psi}_n {\dot \psi}^*_{n'} +\left(m^2 -  \frac{\vec{n}\cdot \vec{n}'}{a^2}\right) \psi_n \psi^*_{n'}\right][A_n^{(0)}, A_{- n'}^{(0)\dagger}]\\
& = & \frac{1}{4 \pi^2} \int_0^\infty n^2 d n \, (|{\dot \psi}_n|^2 + \omega_n^2 |\psi_n|^2),
\end{eqnarray}
where commutation relations \eqref{normalcom} have been used in the second line. From \eqref{adiabatic} and then \eqref{adiabatic2} we get
\begin{eqnarray} 
(|{\dot \psi}_n|^2 + \omega_n^2 |\psi_n|^2) &=& \frac{1}{2 a^3 W_n}\left[\omega_n^2 + W_n^2 +\frac{1}{4} ( \ln a^3 W_n){\dot \ } ^2\right]\\
& = & \frac{1}{a^3 \omega_n}\left[\omega_n^2 + \frac{1}{8} \left(3 H + \frac{{\dot \omega}_n}{\omega_n}\right)^2 + {\cal O} (n^{-2})\right]. \nonumber
\end{eqnarray}
Note that $\epsilon_2$ drops from the expression of the energy momentum tensor VEV if we do not consider beyond the quadratic divergence. In other words, all we need of  $W_n$  is its expression at leading adiabatic order, $W_n \sim \omega_n$. We finally get
\begin{equation} \label{square}
\langle 0 |\, T_0^0 (t, \vec{x}) \, | 0 \rangle = \frac{1}{4 \pi^2 a^3} 
\int_0^\infty n^2 d n \, \left[\omega_n + \frac{H^2 a }{2 n} + {\cal O}(n^{-3})\right],
\end{equation}
to be opposed to the flat space result
\begin{equation} \label{square2}
\langle 0 |\, T_0^0 (t, \vec{x}) \, | 0 \rangle_{\rm flat} = \frac{1}{4 \pi^2 a^3} 
\int_0^\infty n^2 d n \, \omega_n .
\end{equation}

\section{Getting rid of Quadratic Divergences in a toy-model Universe} \label{sec4.5}

In order to enforce the Ultra Strong Equivalence Principle we want to get rid of the terms in the local energy momentum VEV that we normally would not find in flat space. In this paper we deal  only with the time dependent quadratic divergence, the second term inside the square brackets in \eqref{square}.

First of all, we focus our attention on a given comoving observer/trajectory at any proper time $t$ and, say, at point $\vec{x} = 0$. For brevity, when we write $\vec{x} \approx 0$, we refer to a region of space around that trajectory small enough that the manifold description holds and we can define spatial gradients and local commutation relations between neighboring points.

We want to 
strictly preserve the local dynamics of the scalar field. Our starting points will be, therefore, the 
field equations \eqref{equation} in the Heisenberg picture for the field operator in $\vec{x} = 0$,
\begin{equation} \label{evo}
\ddot \phi(t, \vec{x} \approx 0) + 3 H \dot \phi (t, \vec{x} \approx 0) + \left(m^2 - \frac{\partial_i^2}{a^2}\right)  \phi (t, \vec{x} \approx 0)= 0
\end{equation}
and the Hamiltonian density ${\cal H}(t, \vec{x}\approx 0)$ at point 0, which, again, is an explicitely time dependent operator:
\begin{equation} \label{hamtime}
T^0_0 =  {\cal H}(t, \vec{x} \approx 0) = \frac{1}{2}\left[ \dot{\phi}^2 (t, \vec{x} \approx 0) + \frac{\partial_i \phi^2 (t, \vec{x} \approx 0)}{a^2} + m^2 \phi^2 (t, \vec{x} \approx 0) \right] .
\end{equation}
While \eqref{equation} and \eqref{calH} are both elegantly derivable from the same action, for the moment we are forced to consider \eqref{evo} and \eqref{hamtime} as two basic -- and in principle independent -- ingredients of this novel framework. Note moreover that, in a sort of pre-geometric fashion,
we can take the above equations as the implicit definition of the local expansion $a(t)$. By homogeneity, we will require in Sec. \ref{secglobal} that the translated fields at some other point in space still obey \eqref{evo} with the same scale factor $a(t)$.
Commutation relations between local quantities are strictly preserved:
\begin{equation} \label{local}
[\phi (\vec{x}), \pi(\vec{x}' \approx \vec{x})] = i\delta^3(\vec{x}-\vec{x}'),
\end{equation}
(the momentum conjugate to $\phi$ can still be defined completely locally in terms of the Lagrangian density in $\vec{x} = 0$: $\pi(\vec{x}\approx 0) = \partial {\cal L}(\vec{x}\approx 0)/\partial \dot \phi (\vec{x}\approx 0)$).

As before, we formally expand the field in Fourier modes:
\begin{equation} \label{field2}
\phi(t, \vec{x}\approx 0) = \frac{1}{(2 \pi)^{3/2}}\int d n^3\left[\psi_n(t) e^{i \vec{n} \cdot \vec{x}} A^{(1)}_{\vec{n}} + \psi^*_n(t) e^{-i \vec{n} \cdot \vec{x}} A^{(1) \dagger}_{\vec{n}}\right],
\end{equation}
where the suffix $(1)$ means that we now want to consider the modes corrected at first order in the IR expansion. So far the field is only defined at $\vec{x} =0$; the $\vec{x}$ labels at the exponent in the equation above are there just to remind that, if we take space derivatives of the such a local field, a factor $i \, \vec{n}$ drops. The crucial observation now is that the commutator between ${A}$ and ${A}^\dagger$ is proportional to the total volume and therefore will receive the postulated corrections \eqref{2}. By looking at \eqref{bahbah} and \eqref{square}, it is immediate to note that the quadratic divergence is reabsorbed if we impose the commutation relation 
\begin{equation}
[A_{\vec{n}}^{(1)}, A_{\vec{n}'}^{(1)\, \dagger}] = \delta^3(\vec{n} - \vec{n}')\left(1 - \frac{H^2 a^2}{2 n^2}\right).
\end{equation}
The correction is in fact of the expected form \eqref{2}. This is perhaps more evident in the  approach followed in \cite{ultra} where a compact toy-Universe of total linear size $L$ much smaller than the Hubble length was choosen to begin with. In that framework, Fourier labels assume discrete values and the correction $H^2 a^2/n^2$ is clearly of order $H^2 L^2$ or smaller.   Although we previously referred to modifications at scales of the inverse curvature, we find that, more precisely, it is the extrinsic curvature which appears to regulate the correct modification. Equivalently, we can just say that the annihilators are modified with respect to GR by
\begin{equation}
A_{\vec{n}}^{(1)} = \sqrt{1 - \frac{H^2 a^2}{2 n^2}} A_{\vec{n}}^{(0)} .
\end{equation}

The corrected Fourier modes are effectively time dependent. At least at this order of approximation, however, the time dependence is just a multiplicative factor that leaves the vacuum untouched. Note also that, by applying the postulated local equation \eqref{evo}, the time dependence of $A^{(1)}_n$ introduces additional spurious terms inside \eqref{modes0}. Those, however, are of higher order in the $n$ expansion and therefore do not affect the present result. Finally, inverse Fourier transforms (such as inverting \eqref{field2} by integrating over $\vec{x}$) are not allowed. Following the reasoning of section \ref{sec3}, we argue that whenever we take integrals $\int d^3 x$ of operators over a volume $V$ we are making an error of order ${\cal O}(H^2 V^{2/3})$.

So far we concentrated only on local operators at some given point $\vec{x}=0$ and made connection with the global Fourier modes. A crucial ingredient to understand the large-scales 
properties of this deformed spacetime is the momentum operator constructed with the modified Fourier modes,
\begin{equation} \label{momentum1}
\vec{P}^{(1)} = \int d^3 n\, \vec{n} A_{\vec{n}}^{(1)\dagger} A_{\vec{n}}^{(1)} .
\end{equation}
In fact, we are not provided with a metric element that we can integrate over, but we can exploit the assumed homogeneity of this spacetime and define a translation operator by simply exponentiating 
\eqref{momentum1}. By applying the translation operator, in section \ref{secglobal} we are able to define the local operators a finite distance away from $\vec{x}=0$. 

\section{Refining the Model} \label{sec5}

The form of the momentum operator suggests another way to look at the proposed deformation. 
We can write \eqref{momentum1} in terms of the usual GR operators $A_{\vec{n}}^{(0)}$, satisfying 
the usual commutation relations, and modify the momentum labels instead:
\begin{equation} 
\vec{P}^{(1)} = \int d^3 n\, \vec{k} \, A_{\vec{n}}^{(0)\dagger} A_{\vec{n}}^{(0)} ,
\end{equation}
where, at this order of approximation, 
\begin{equation} \label{kk2}
\vec{k} = \vec{n} \left(1 - \frac{H^2 a^2}{2 n^2} \right). 
\end{equation}
We are trading modified Fourier operators for modified dispersion 
relations. In other words, in this equivalent, but more practical, formulation, we are postulating a slight mismatch between the ``metric-manifold" labels $\vec{n}$ and the physical momenta $\vec{k}$ that locally define the infinitesimal translations and the derivatives of the local fields.
Since $\vec{n}$ is the usual comoving ``manifold" label, we assume that it is time-independent. Moreover, since we will use only the usual ``manifold" creators operators $A_{\vec{n}}^{(0)}$ from now on, we drop the  the suffix $(0)$ to simplify the notation. 
The recipe can finally be summarized as follows:

\begin{itemize}
\item There are two different (comoving) Fourier coordinates: the ``manifold"-coordinate $\vec{n}$ and the locally defined physical momentum $\vec{k}$. The integration measure in Fourier space is flat in the $n$-coordinates. Moreover, during time evolution, $\vec{n} $ is conserved, $\vec{k}$ is not. The relation between the two is
\begin{equation} \label{kk}
\vec{k} = \vec{n} \left(1 - \frac{H^2 a^2}{2 n^2} + {\rm higher\ order} \right).
\end{equation}

\item The local field in $\vec{x}\approx 0$ is expanded in Fourier modes as follows:
\begin{equation} \label{localfield}
\phi(t, \vec{x}\approx 0) = \frac{1}{(2 \pi)^{3/2}}\int d n^3\left[\psi_k(t) e^{i \vec{k} \cdot \vec{x}} A_{\vec{n}} + \psi^*_k(t) e^{-i \vec{k} \cdot \vec{x}} A^\dagger_{\vec{n}}\right] \equiv \frac{1}{(2 \pi)^{3/2}}\int d n^3 \phi_{\vec k}(t) e^{i \vec{k} \cdot \vec{x}} 
\end{equation}
and the $A_n$ satisfy the usual commutation relations: 
\begin{equation} \label{usualcom}
[A_{\vec{n}}, A^\dagger_{\vec{n}'}] = \delta^3(\vec{n} -\vec{n}').
\end{equation}

Again, the $\vec{x}$ coordinate in \eqref{localfield} extends only as far as is needed to take the space derivatives of the field in $\vec{x}=0$. When a derivative is taken, a factor of $\vec{k}$, instead of $\vec{n}$, drops. 

\item The local equation for $\phi(t,\vec{x}\approx 0)$, \eqref{evo}, fixes the equations for the functions $\psi_n$:
\begin{equation}
\ddot{\psi}_k(t)  + 3 H \dot{\psi}_k(t) +  \omega^2_k \psi_k(t) = 0, 
\end{equation}
where
\begin{equation}
\omega_k = \sqrt{\frac{k^2}{a^2} + m^2} 
\end{equation}
and $\vec{k}$ is given in \eqref{kk}. The solution of the above equations are choosen to correspond to the adiabatic vacuum (in the physical momentum $k$), i.e.
\begin{equation} \label{adiabatic3} 
\psi_k(t) = \frac{1}{\sqrt{2 a^3 W_k}} \exp \left(-i \int^t W_k dt'\right).
\end{equation}
The function $W_k$ can be approximated at arbitrary adiabatic order as before. The only difference is that $\omega_k$ should substitute $\omega_n$ in \eqref{adiabatic2} and \eqref{eps2}.

\item The energy density in $ \vec{x} \approx 0$ is expressed in terms of the local field $\phi(t,\vec{x}\approx 0)$ in the usual way (equation \eqref{hamtime}), 
\begin{equation} \label{energymomentum}
T^0_0 =  {\cal H}(t, \vec{x} \approx 0) = \frac{1}{2}\left[ \dot{\phi}^2 (t, \vec{x} \approx 0) + \frac{\partial_i \phi^2 (t, \vec{x} \approx 0) }{a^2} + m^2 \phi^2 (t, \vec{x} \approx 0) \right] .
\end{equation}

\end{itemize}

The above equations can be used to calculate the vacuum expectation value of $T_0^0$. By direct substitution of \eqref{localfield} into \eqref{energymomentum} we get
\begin{eqnarray} \nonumber
\langle 0 |\, T_0^0 (t, \vec{x}\approx 0) \, | 0 \rangle & = & \frac{1}{2(2\pi)^3}
\int d^3 n \, d^3 n' \left[{\dot \psi}_k {\dot \psi}^*_{k'} +\left(m^2 -  \frac{\vec{k}\cdot \vec{k}'}{a^2}\right) \psi_k \psi^*_{k'}\right][A_n, A_{- n'}^{\dagger}]\\
& = & \frac{1}{4 \pi^2} \int_0^\infty n^2 d n \, (|{\dot \psi}_k|^2 + \omega_k^2 |\psi_k|^2),
\end{eqnarray}
By substitution of \eqref{adiabatic3} we then get
\begin{eqnarray} 
\langle 0 |\, T_0^0 (t, \vec{x}\approx 0) \, | 0 \rangle &=& \frac{1}{4 \pi^2 a^3} 
\int_0^\infty n^2 d n \, \left[\omega_k + \frac{H^2 a }{2 k} + {\cal O}(n^{-3})\right], \\
& = & \frac{1}{4 \pi^2 a^3} 
\int_0^\infty n^2 d n \, \left[\omega_n  + {\cal O}(n^{-3})\right]
\end{eqnarray}

Note that the quadratic divergences are here reabsorbed just be re-expressing $\langle 0 |\, T_0^0  | 0 \rangle$ in terms of the appropriate ``flat measure" time-independent Fourier coordinates $n$.

\section{A look at the global picture} \label{secglobal}

So far, we have defined local operators only in $\vec{x}\approx 0$ and made connection with the global ones. However, we want to be able to move in this space arbitrarily far in any direction and define local operators also elsewhere. 
We are provided with a momentum operator,
\begin{equation} \label{momentum}
P_i =  \int d^3n\, \vec{k}(\vec{n})\,  A^\dagger_{\vec{n}} A_{\vec{n}}\, 
\end{equation}
where, again,
\begin{equation}\label{kkk}
\vec{k}(\vec{n}) = \vec{n} \left(1 - \frac{H^2 a^2}{2 n^2}\right).
\end{equation}
Since we are assuming homogeneity, we obtain a global translation operator by exponentiation,
\begin{equation}T_i(\lambda) = e^{-i \lambda P_i} . \label{translation}
\end{equation}
Thus, we can define local operators at a finite distance from $\vec{x}\approx 0$ by the action of $T_i$:
\begin{equation} \label{keepgoing}
\phi(t, *) \equiv
T_i(\lambda)\, \phi(t, 0) \, T_i^{-1}(\lambda) =  \frac{1}{(2 \pi)^{3/2}} 
 \int d^3 n\, \phi_{\vec{k}} \ e^{- i\, \lambda \, n_i \left(1- \frac{H^2 a^2}{2 n^2}\right)}.
\end{equation}
where $*$ stands for ``at distance $\lambda$ if one keeps going in the $i$ direction" and $\lambda$ is made up by all the infinitesimal steps that have been taken in order to go from one point to another and is therefore the (comoving) proper distance. The momentum operator is given in \eqref{momentum}-\eqref{kkk} only approximately. Therefore, also \eqref{keepgoing} should be trusted only as far as $d = a \lambda$ is smaller than the Hubble radius.

In the present case, because of homogeneity, 
translations in different directions commute with each other, so we can label the local operators with three parameters and describe, approximately, a finite patch of Universe around $\vec{x}\approx 0$. Explicitely,

\begin{equation} \label{localfar}
\phi(t, \vec{\lambda}) =  \frac{1}{(2 \pi)^{3/2}} 
 \int d^3n\,  \left[\psi_k(t) e^{i \vec{k} \cdot \vec{\lambda}} A_{\vec{n}} + \psi^*_k(t) e^{-i \vec{k} \cdot \vec{\lambda}} A^\dagger_{\vec{n}}\right].
\end{equation}
Although the above equation has a familiar form, the labels $\lambda^i$ should really be thought here as the parameters of the translation group. We will soon show that, if we use them as integration variables at $t =$ const., as we would do on a standard FRW flat space, we make an error of order $V^{2/3} H^2$.

It is interesting to calculate, at some given time $t$, the canonical commutator between $\pi(0) = a^3 \dot \phi(0)$ and the field itself at a distance. By using the standard commutation relations 
\eqref{usualcom} we obtain
\begin{equation}
[\pi(0), \phi( \vec{\lambda})] = - \frac{a^3}{(2\pi)^3} \int d^3 n \, \left[\dot{\psi}_k^* \psi_k e^{i \vec{k} \cdot \vec{\lambda}} - \dot{\psi}_k \psi_k^* e^{- i \vec{k} \cdot \vec{\lambda}}\right] .
\end{equation}
By using the Wronskian condition $a^3  (\dot{\psi}_k^* \psi_k - \dot{\psi}_k \psi_k^*) = i $, immediately derivable from \eqref{adiabatic3},  and \eqref{kkk} we get, up to higher orders in $H a \lambda$,
\begin{equation}
[\pi(0), \phi( \vec{\lambda})] = -\frac{i}{(2\pi)^3} \int d^3 n\, e^{- i\, \vec{\lambda}\cdot \vec{n} \left(1- \frac{H^2 a^2}{2 n^2}\right)} \simeq 
-\frac{i}{(2\pi)^3} \int d^3 n\, e^{- i\, \vec{\lambda}\cdot \vec{n}} \left(1+ i \vec{\lambda}\cdot \vec{n} \frac{H^2 a^2}{2 n^2}\right),
\end{equation} 
where we have expanded the exponential in the last expression. The first term gives the usual delta function. The last term gives
\begin{equation}
\frac{i H^2 a^2}{2 (2\pi)^3} \frac{d}{d\alpha}\left.\int \frac{d^3 n}{n^2} e^{- i\, \alpha\, \vec{\lambda}\cdot \vec{n}} \right|_{\alpha = 1}\ =\ - \frac{i}{8 \pi} \frac{H^2 a^2}{\lambda}\, .
\end{equation}
In this way we obtain the first correction to the canonical commutator, 
\begin{equation} \label{comm}
[\pi(0), \phi( \vec{\lambda})] = -i\left(\delta^3(\vec{\lambda}) + \frac{1}{8 \pi} \frac{H^2 a^2}{\lambda} \right).
\end{equation}
The above relation finally gives a more precise sense to \eqref{local}. At face value, \eqref{comm} is a signal of a non-locality that becomes more and more important at large scales. But again, that is the interpretation that we would have of \eqref{comm} on a metric manifold. 
In section \ref{sec3} we argued that regions of space at a given time can still be associated with a well defined algebra of ``averaged" operators of the matter theory, but that those operators can only approximately be expressed as integrals over a metric manifold. The expression in \eqref{comm} is in fact a three-dimensional density and the correction term is just signaling the mistake that we make if we naively take the integral 
\begin{equation}
\int_{\lambda a\ll H^{-1}}  d^3 \lambda'\, [\pi(0), \phi(\vec{\lambda}')] \ =\  -i \left[1 + \frac{1}{4} H^2 a^2 \lambda^2 +{\cal O}(H^2 V^{2/3})^2 \right]
\end{equation}
and use $\lambda$ as a flat-measure integration variable.
We nicely recover the expected pattern \eqref{2}.

It is equally interesting to calculate $[\phi(0), \pi(\lambda)]$. First, we note that there is a potential ambiguity in defining the time derivative of a displaced operator. 
By deriving \eqref{localfar} we get
\begin{eqnarray} 
 \pi(\vec{\lambda}) &=&  \frac{a^3}{(2 \pi)^{3/2}} \nonumber
 \int d^3n\,  \left[\dot{\psi}_k(t) e^{i \vec{k} \cdot \vec{\lambda}} A_{\vec{n}} + \dot{\psi}^*_k(t) e^{-i \vec{k} \cdot \vec{\lambda}} A^\dagger_{\vec{n}}\right] \\
 &+& i \frac{a^3}{(2 \pi)^{3/2}}  \int d^3n\,  (\vec{k} \cdot \vec{\lambda})\dot{}\ \left[\psi_k  \ e^{i \vec{k} \cdot \vec{\lambda}} A_{\vec{n}}  - \psi^*_k  \ e^{- i \vec{k} \cdot \vec{\lambda}} A^\dagger_{\vec{n}} \right] . \label{ambig}
\end{eqnarray}
The second line in the above equation is there because $k$ is time dependent. 
However, if we just applied the translation to $\pi(0)$, instead of deriving $\phi(\vec{\lambda})$, those terms would not be there. Therefore, for consistency, we need to make them ineffective at the required order of approximation. In other words, we have to impose that  $[\phi(0), \pi(\vec{\lambda})] = - [\pi(0),\phi(\vec{\lambda})]$. Because of the second line of \eqref{ambig}, the commutator between $\phi(0)$ and $\pi(\lambda)$ gives
\begin{equation} \label{nonsym}
[\phi(0), \pi(\vec{\lambda})] = - [\pi(0),\phi(\vec{\lambda})] - 2i \frac{a^3}{(2\pi)^3}
\int d^3 n \, e^{- i \vec{k} \cdot \vec{\lambda}} \left| \psi_n \right|^2 (\vec{k} \cdot \vec{\lambda})\dot{} \ ,
\end{equation}
the last term being the spurious contribution. In order to get rid of it, we have to  make the comoving physical distance $\vec{\lambda}$ also time dependent. This effectively means that, after an infinitesimal time step $dt$, we have to reconsider the field translated, from  $\vec{x} \approx 0$, not by the same comoving distance $\lambda$, but by a slightly different amount. 
At high momenta/small distances, since $|\psi_k|^2 \sim 1/n$, the integral in the second term of \eqref{nonsym} reads
\begin{equation}
\int d^3 n \, e^{- i \vec{n}  \cdot \vec{\lambda}} \, \frac{1}{n} \left[\dot{\vec{\lambda}} \cdot \vec{n}\left(1 - \frac{H^2 a^2}{2 n^2}\right) - \vec{\lambda}\cdot \vec{n} \frac{(H^2 a^2)\dot{}}{2 n^2}\right]\, .
\end{equation}
We make the ansatz $\dot{\vec{\lambda}} = \beta \lambda^2 (H^2 a^2)\dot{}\, \vec{\lambda}$, $\beta$ being a number to be determined. We get
\begin{equation}
\int d^3 n \, e^{- i \vec{k} \cdot \vec{\lambda}} \left| \psi_n \right|^2 (\vec{k} \cdot \vec{\lambda})\dot{}\ \simeq \ i  (H^2 a^2)\dot{} \ \frac{ d}{ d \alpha} \int d^3 n\, \left. \left(\frac{\beta \lambda^2}{n} - \frac{1}{2 n^3}\right) e^{- i \alpha\, \vec{n} \cdot \vec{\lambda}}\right|_{\alpha =1}.
\end{equation}
The last integral can be regularized by setting $n^{-3} \rightarrow n^{-3 + \epsilon}$ and taking the $\epsilon \rightarrow 0$ limit only after deriving with respect to $\alpha$. The result is null for $\beta = 1/4$, which fixes the time dependence of $\lambda$:
\begin{equation} \label{expansion}
\dot{\lambda} = \lambda^3 \frac{(H^2 a^2)\dot{}}{4} .
\end{equation}

The above formula is one of the main results of this paper and can be rephrased as follows. In this homogeneous Universe, in the vicinity of each point, there is a local expansion given by the same scale factor $a(t)$. In essence, $a(t)$ rules the evolution equations for the local operators and determines the rate at which  the distance between two comoving observers grows, as far as such a distance is extremely small compared to Hubble. In that limit, the comoving distance $\lambda$ is conserved.  
However, if you pick up a pair of comoving observers further apart, their distance 
grows slightly differently, which means that the expansion rate on large scales is effectively dependent on the distance. The modified expansion rate is given, at first order, by \eqref{expansion}. This effect, which is the most striking signal of the breakdown of the manifold description, is clearly negligible well within the Hubble scale.

\section{Cosmology}\label{seccosm}

Cosmology is where the predicted departures from GR are more relevant, since every high red-shift object is located from us at distances comparable to the inverse curvature. 
Eq. \eqref{expansion} can be integrate straightforwardly:
\begin{equation}\label{lambda}
\lambda^{-2}(t) - \lambda^{-2}(t') = -\frac{1}{2}[H^2(t) a^2(t) - H^2(t') a^2(t')].
\end{equation}
Equivalently, the proper distance $d = a \lambda$ between two comoving observers evolves from time $t'$ to $t$  as
\begin{equation} \label{d}
\frac{d(t)}{d(t')} = \frac{a(t)}{a(t')}\left[1 - \frac{d^2(t')}{2}\left(H^2(t) \frac{a^2(t)}{a^2(t')} - H^2(t')\right)\right]^{-1/2}.
\end{equation}
According to \eqref{lambda} and \eqref{d} the distance between two comoving observers grows slightly less than in a FRW Universe with scale factor $a(t)$.

We now want to calculate the ``corrected" trajectory $r(\tau)$ for a light ray, where $r$ is its comoving distance and $\tau$ the conformal time. 
A light ray passing through $\vec{x}\approx 0$ satisfies $d r = d\tau$. However, on top of the usual contribution, the trajectory of a light ray receives a correction from the modified global expansion \eqref{expansion}:
\begin{equation}
\frac{d r}{d \tau} = 1 + r^3 \frac{(H^2 a^2)'}{4} .
\end{equation}
Now we specialize to a matter dominated Universe for simplicity.  Therefore, we set $H a = 2/\tau$ and we get
\begin{equation} \label{22}
\frac{d r}{d\tau} = 1 - 2\frac{r^3}{\tau^3}.
\end{equation}
Normally, at the RHS we would simply have $1$ and, again, the second term should be considered just as the first order correction.
We couldn't find an analytic solution to the above equation, and therefore we will solve it numerically. With respect to an Einstein-De Sitter Universe, a light ray propagating forward in time will go a shorter way. A light ray that we are detecting now, on the other hand, is actually coming from farer.  For a light ray propagating back in the past the above equation becomes, in terms of the redshift,
\begin{equation} \label{backintime}
\frac{d (H_0 r)}{dz} = \frac{1}{(1+z)^{3/2}} +  \frac{(H_0 r)^3}{4}.
\end{equation}
The above should be solved with initial condition $r(0) = 0$ and will give some $r(z)$ to be used in what follows. Note that, as in the normal framework, the Hubble parameter $H_0$ is just a total normalization factor for cosmological distances. 

Light rays going backward in time in any direction describe, at some redshift $z$, the surface of a two-sphere of area $4 \pi a(z)^2 r(z)^2$. Therefore, 
the angular diameter distance is $r(z)$ multiplied by $a(z)$:
\begin{equation}
d_A(z) = \frac{1}{1+z} r(z).
\end{equation}
The redshift of the frequences, on the other hand, is basically unmodified in this framework, $1+z = 1/a$. It is true that the comoving momentum $\vec{k}$ \eqref{kkk} is not strictly conserved, but the effect is negligible for momenta much bigger than $H$. 

\begin{figure}[tbp]
  \begin{center}
    \subfigure[Present Framework]{\label{usep}\includegraphics[width=3.3in]{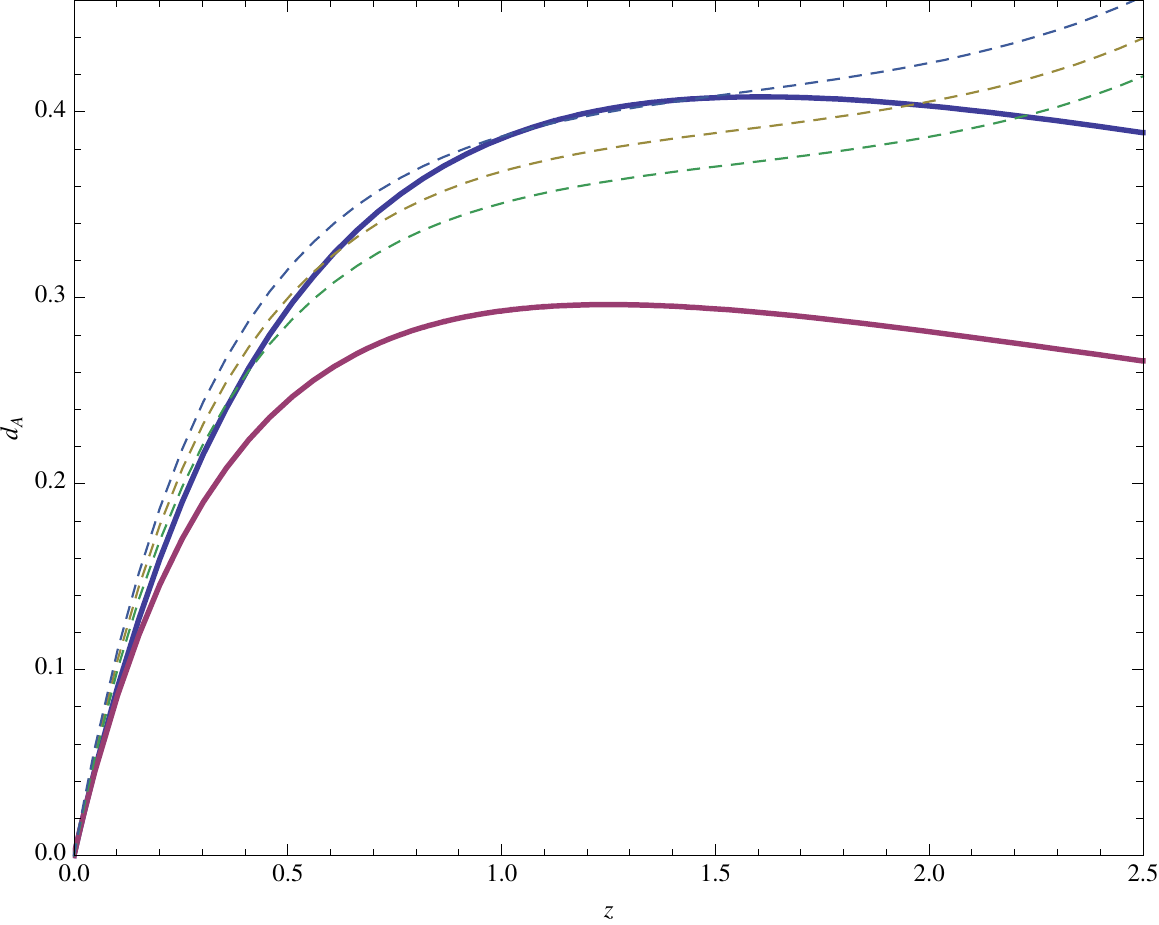}}
    \subfigure[Einstein-de Sitter]{\label{edesit}\includegraphics[width=3.3in]{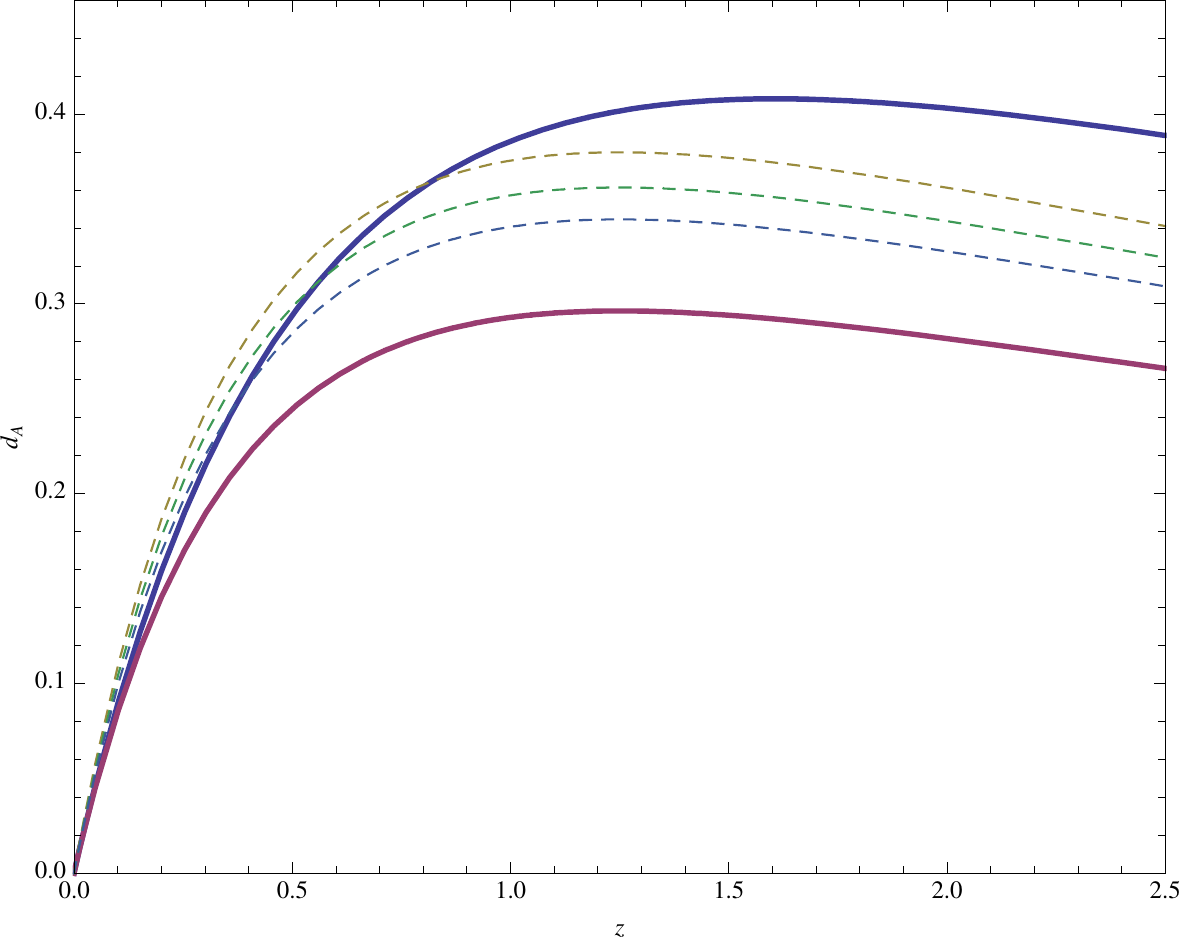}} \\
  \end{center}
\caption{In both figures the solid curves show the angular diameter distance $d_A$ for $\Lambda$CDM model with $\Omega_m = 0.3$ (upper curve) and for  Einstein-de Sitter  ($\Omega_m = 1$, lower curve) where the Hubble constant $H_0$ has been normalized to 1. In the left panel we compare that behaviour with what we find in the present model. In order to obtain curves similar to $\Lambda$CDM we have to lower the Hubble constant $H_0$ which is the only parameter left free in this model.  From top to bottom we show the results for $H_0 = 0.78, 0.82, 0.86$. The runaway behaviour after $z\simeq2$ is probably due to the fact that we are getting very close to the Hubble radius at that redshift (we find, at $z=2$, $d_A \simeq (2/3) H^{-1}$) and therefore higher order IR corrections to our formulas are needed. The right panel shows the effect of lowering the Hubble constant to those same values in the unmodified matter dominated Einstein-de Sitter Universe.} 
  \label{fig1}
\end{figure}

\begin{figure}[tbp]
  \begin{center}
    \subfigure[Present Framework]{\label{usep2}\includegraphics[width=3.3in]{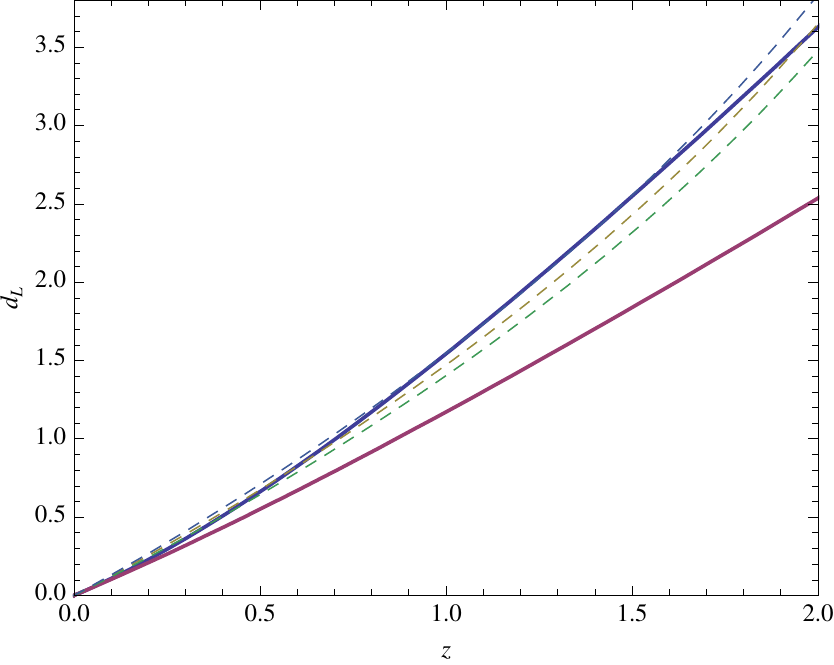}}
    \subfigure[Einstein-de Sitter]{\label{edesit2}\includegraphics[width=3.3in]{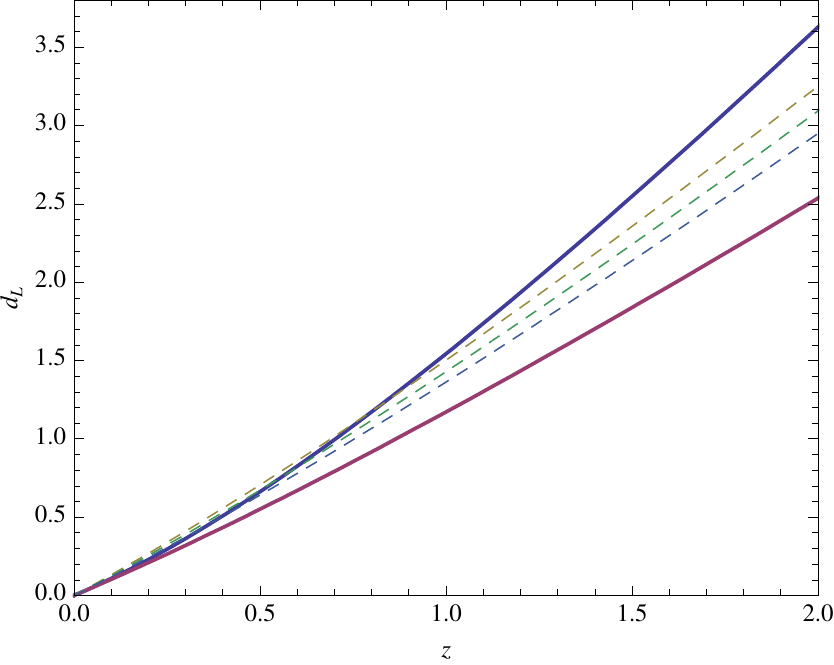}} \\
  \end{center}
\caption{Same as Fig. \ref{fig1} but for the luminosity distance $d_L = (1+z)^2 d_A$.} 
  \label{fig2}
\end{figure}

Obtaining the formula for the luminosity distance looks, in the present framework, slightly less straightforward. We discuss its derivation in the Appendix where we argue that, at least of this order of approximation, angular diameter and luminosity distances are related by the usual formula 
\begin{equation}
d_L = (1+z)^2 d_A\, .
\end{equation}
We plot the obtained luminosity distance as a function of the redshift in Fig. \ref{fig2} for different values of the Hubble parameter.

Our correction makes cosmological distances larger than in a usual Einstein-de Sitter Universe. This goes in the right direction of a $\Lambda$CDM Universe, at least qualitatively (Figs. \ref{fig1} and \ref{fig2}, left panel). However, this effect kicks in only at a relatively high redshift. By expanding $r(z)$ as $r(z) = r_1 z + r_2 z^2+\dots$ it is easy to see that the proposed effect only modifies the curve at forth order in the small-z expansion, while dark energy models usually give already contributions at second order.  Qualitatively, in order to raise the overall angular diameter distance to the plateau reached by $\Lambda$CDM, we have to lower the Hubble constant $H_0$, so that $d_A(z)$ starts out with a slightly higher derivative. This makes our curves systematically higher than $\Lambda$CDM at small redshift. Note, on the other hand, that lowering $H_0$ in the absence of this effect produces curves of a completely different shape (Figs. \ref{fig1} and \ref{fig2}, right panel). Finally, the runaway behaviour of $d_A$ after about $z=2$ (Fig. \ref{fig1}, left panel) is probably due to the fact that the distance covered by the light ray becomes dangerously comparable to Hubble at that redshift (indeed, $d_A(z=2) \simeq (2/3) H^{-1}(z=2)$) and therefore higher order corrections to the present model are needed.

While the low-redshift discrepancies with the successful concordance model suggest that this mechanism might not be able, by itself, to give full account for the apparent acceleration of the Universe (for a more quantitative analysis see \cite{savvas}), it is worth noting that the case for an Einstein-de Sitter Universe with a lower Hubble constant has occasionally been made in the literature. Interestingly, the authors of \cite{sarkar} claim that a model with $\Omega_m = 1$ and $H_0 = 46$ km/s/Mpc fits cosmological observations, with the exception of supernovae, better than the concordance model.

\section{Discussion} \label{sec9}

Within the established low-energy effective framework ``GR + matter fields" 
\begin{equation} \label{act}
S = \int \sqrt{-g}\left(M_{\rm Pl}^2 R + {\cal L}_{\rm matter}\right)
\end{equation}
there are models that can fit our cosmological data with impressive accuracy. However,  those explanations do not come without a cost: we need to assume two epochs of accelerating expansion (inflation and ``dark energy"), characterized by widely separated  mass scales, that require appropriate negative pressure components and, expecially for dark energy, a tremendous amount of fine tuning. With the possible exception of inflation, 
these difficulties look oblivious to the possible UV-completions of \eqref{act}. The candidate theories of quantum gravity, as long as they reproduce \eqref{act} at low-energy, are not expected to essentially simplify the picture, although they proved able to suggest new mechanisms, give insights on the cosmological singularity, motivate the introduction of new degrees of freedom/sectors needed for the acceleration etc\dots

In this paper we questioned the IR-side of the paradigm ``GR + matter fields" and attempted to look at \eqref{act} as a small distance approximation. The proposed modification is radical because it challenges the usual geometrical description of spacetime on the largest scale, but compelling because it does not contain any adjustable parameter: in a cosmological set-up deviations from GR are calculable and become relevant, at any epoch, at the Hubble scale. Abandoning the well understood and reliable framework \eqref{act} is adventurous to say the least; but we tried to do that by following a physical principle that seems to address, at least in part, some theoretical difficulties such as the cosmological constant problem and the black hole information paradox and, at the same time, to give observable cosmological predictions at high redshift. 

The most relevant cosmological effect is a departure at large scales from the local expansion rate (e.g. eq. \eqref{exintros}). 
The general picture that this model suggests is that the present acceleration of the Universe may not be due to any additional ``dark energy" component but to a systematic  breakdown of the geometrical description of spacetime at length scales comparable to Hubble. The proposed effect does not kick in at some given time ``by coincidence" during the cosmological evolution, but is permanently present at any cosmological epoch whenever a comoving observer looks at very far away objects. 

Exploring new directions often opens up potential problems and questions which have been answered here only in part and surely need future work and further double checks. We isolated the IR corrections that apply to a free scalar field in a spatially flat FRW Universe. It is not clear if those corrections have some sort of universality i.e. if they automatically apply to, or are inherited by, other types of fields. For instance, in a cosmological set-up, applying the Ultra-Strong Equivalence Principle to the electromagnetic field is just trivial, because of conformal invariance. For photons, the time dependent terms that we got rid of in the high momentum expansion~\eqref{structure} are absent from the very beginning. If the proposed model is really a modification of the underlying geometry, such a modification should be universal, and this, at present, is not guaranteed. 

It would be interesting to consider the implications of this model in the very early Universe and close to the singularity, where IR corrections of order Hubble might become important. 
We note also that the usual cosmological kinematic of the modes (``exiting" during acceleration, ``entering" during deceleration) is modified in the present framework, as the physical comoving momentum is not conserved on large scales (eq. \ref{conserved}). It would be interesting to see what are the implications of this effect in the early Universe, although higher orders in the IR expansion might be needed in order to have a more complete picture.

Whether or not the proposed modification applies to reality, our findings are of some interest and can be rephrased as follows:
how much should we modify GR  in order to ``renormalize" the stress tensor \eqref{structure} from the IR side -- with an IR modification -- instead than with local gravitational counterterms? How different from GR would the conjectured theory be? We have shown that the required modification is indeed extremely tiny. For a free scalar in a cosmological set-up, Fourier modes of physical momentum ${\vec p}$ get modified by an amount of order $H^2/p^2$. On cosmological scales, such a modification goes in the direction of an effective acceleration but is still too weak to give full account of supernovae observation~\cite{savvas}. It will also be interesting to modify, along these same lines, a black hole or a stellar solution, also to check the compatibility with constraints coming from the solar system and strongly self gravitating objects. At the level of order of magnitude estimate, however, we do not expect relevant corrections. The proposed modification affects those phenomena whose relevant length scales are of order the inverse curvature. Atomic spectra and and nuclear physics should be untouched, even in the presence of strong gravitational fields. The reason is that the inverse (Ricci) curvature is roughly of order $M_{\rm Pl}^2 /\rho$, $\rho$ being the energy density, and this typically gives a huge number in atomic size units. For what concerns the solar system, its entire dynamics is well within the average curvature radius.
If we try to compare the typical solar system distances to, say, the Weyl curvature, those are always extremely small. An order of magnitude estimate gives $({\rm Weyl\ curvature})^{-1/2} \approx r(r/3{\rm Km})^{1/2}$, $r$ being the distance from the sun. This means that, within the solar system, and as opposed to high-redshift cosmology, objects are always at distances much smaller than the inverse curvature.  

We finally note that some recent observational inputs in cosmology may look as an encouragement for IR modifications of gravity. Beside the acceleration of the Universe, which is already puzzling by itself,  some emerging tensions between standard $\Lambda$CDM cosmology and large-scale observations might be indicating that there is something going wrong with the standard GR description of the largest scales. 
Interestingly, other IR modifications of GR have already been tested \cite{justin, mark} against those discrepancies, which include, among other, the lack of large-scale CMB correlations~\cite{COBE}, the anomalously large
integrated Sachs-Wolfe cross-correlation~\cite{seljak} and the abnormally large bulk flows on very large scales~\cite{kash}.

\vspace{1cm}

{\bf Acknowledgments}

I thank Niayesh Afshordi, Sergio Cacciatori, Fabio Costa, Paolo Creminelli, Olaf Dreyer, Justin Khoury, Jennifer Lin, Christian Marinoni, Savvas Nesseris, Burt Ovrut, Filippo Passerini, Costantinos Skordis, Brian Schmidt, Lee Smolin,  Shinji Tsujikawa, Andrew Tolley and Mark Wyman. Research at the Perimeter Institute is supported in part by the Government of Canada through NSERC and by the Province of Ontario through the Ministry of Research \& Innovation.\\

\appendix

\section{The luminosity distance}

The luminosity distance accounts for the flux of energy per unit time arriving from the source. 
In standard cosmology it is known that there are two in principle-independent dampening cosmological effects for the flux. The usual analysis nicely relies on the wave-particle duality: First, each photon has its energy red-shifted by a factor $1/(1+z)$. Second, the number of them per unit time decreases, from emission to absorption, by the same factor. As already discussed, the present framework does not change the first effect: comoving momenta much higher than Hubble are conserved and therefore physical momenta redshift in the usual way. Here we argue that also the second dampening effect is proportional to $1/(1+z)$, at least at the present order of approximation. In principle, this is not so obvious. We might in fact reason as follows.

Consider a bunch of photons leaving the source at redshift $z$ and propagating towards us. Say that, at a given redshift $z$, the comoving distance between the first photon and the last is $\Delta(z) = r_{\rm first}(z) - r_{\rm last}(z)$, where both $r_{\rm first}(z)$ and $r_{\rm last}(z)$ are solutions of \eqref{backintime}. In the limit in which those two solutions are very close to each other it is easy to estimate $\Delta(z)$. By writing  equation \eqref{22} for $r_{\rm first}(\tau)$ we get
\begin{equation} \label{delta}
r'_{\rm last} + \Delta' = 1 -\frac{2}{\tau^3}(r_{\rm last}^3 + 3 r_{\rm last}^2 \Delta + 3 r_{\rm last} \Delta^2 + \Delta^3).
\end{equation}
We can now use the equation for $r_{\rm last}$ and drop higher orders in $\Delta$. This gives an estimate for a small spread $\Delta(z)$ around a given solution $r(\tau)$,
\begin{equation}
\frac{\Delta'}{\Delta} = -6 \frac{r^2(\tau)}{\tau^3}.
\end{equation}
By reconverting in redshift this gives
\begin{equation}
\Delta(z) = e^{\frac{3 H_0^2}{4}\int_0^z dz' r^2(z')}.
\end{equation}
Numerical integration shows that the above is indeed a very accurate estimate. 
This way of reasoning would suggest that, on top of the usual spreading effect due to the expansion, there is a sort of ``refocussing" effect, due to the fact that two infinitely close solutions of \eqref{backintime} tend to get closer and closer to each other as they evolve forward in time. For faraway objects that is a very relevant brightening effect.

What is not convincing about the above derivation is that it heavily relies on the estimate of a small spread $\Delta$ even when that is calculated very far away from us. We know that the translation operator \eqref{translation} that we are using is very accurate only in shifting objects relatively close to the center $\vec{x} \approx 0$. The error in estimating a small shift $\Delta(z)$ at high redshift can be comparatively very large.

A more convincing way to proceed is simply to note that two parallel light rays, in the limit when they are very close to each other, should feel only the local expansion $a(t)$. At each time step, by homogeneity, we can put one of the two at $\vec{x} \approx 0$ and calculate the position of the other. This amounts to re-set, at each time step, $r_{\rm last} = 0$ in eq. \eqref{delta} and gives, for the comoving spread $\Delta$, an equation identical to \eqref{expansion}. The effect is negligible each time the spread is much smaller than Hubble. We therefore obtain, at least at this order of approximation, the known relation
\begin{equation}
d_L = (1+z)^2 d_A\, .
\end{equation}


\begin{thebibliography}{99}


\bibitem{massive} See, among many others, 
  C.~Deffayet, G.~R.~Dvali and G.~Gabadadze,
  ``Accelerated universe from gravity leaking to extra dimensions,''
  Phys.\ Rev.\  D {\bf 65}, 044023 (2002)
  [arXiv:astro-ph/0105068]; 
  N.~Arkani-Hamed, H.~C.~Cheng, M.~A.~Luty and S.~Mukohyama,
  ``Ghost condensation and a consistent infrared modification of gravity,''
  JHEP {\bf 0405}, 074 (2004)
  [arXiv:hep-th/0312099];
    G.~Dvali, S.~Hofmann and J.~Khoury,
  ``Degravitation of the cosmological constant and graviton width,''
  Phys.\ Rev.\  D {\bf 76}, 084006 (2007)
  [arXiv:hep-th/0703027].

\bibitem{ultra}
F.~Piazza,
  ``Modifying Gravity in the Infra-Red by imposing an 'Ultra-Strong' Equivalence Principle,''
  arXiv:0904.4299 [hep-th].

\bibitem{heat}
  D.~V.~Vassilevich,
  ``Heat kernel expansion: User's manual,''
  Phys.\ Rept.\  {\bf 388}, 279 (2003)
  [arXiv:hep-th/0306138].

\bibitem{connes} 
  A.~Connes,
  \emph{Noncommutative geometry},
  Academic Press, San Diego, CA, 1994.

\bibitem{sw}
  N.~Seiberg and E.~Witten,
  ``String theory and noncommutative geometry,''
  JHEP {\bf 9909}, 032 (1999)
  [arXiv:hep-th/9908142].



 \bibitem{wein}
  S.~Weinberg,
  ``The cosmological constant problem,''
  Rev.\ Mod.\ Phys.\  {\bf 61}, 1 (1989).

\bibitem{haw}
  S.~W.~Hawking,
  ``Breakdown Of Predictability In Gravitational Collapse,''
  Phys.\ Rev.\  D {\bf 14}, 2460 (1976).

\bibitem{fulling2}
  Y.~B.~Zeldovich and A.~A.~Starobinsky,
  ``Particle production and vacuum polarization in an anisotropic gravitational
  field,''
  Sov.\ Phys.\ JETP {\bf 34}, 1159 (1972)
  [Zh.\ Eksp.\ Teor.\ Fiz.\  {\bf 61}, 2161 (1971)];
  S.~A.~Fulling and L.~Parker,
  ``Renormalization in the theory of a quantized scalar field interacting with
  a robertson-walker spacetime,''
  Annals Phys.\  {\bf 87}, 176 (1974).

 \bibitem{birrell}
  N.~D.~Birrell and P.~C.~W.~Davies,
  ``Quantum Fields In Curved Space,''
{\it  Cambridge, Uk: Univ. Pr. ( 1982) 340p}

\bibitem{savvas}
 S.~Nesseris, F.~Piazza and S.~Tsujikawa,
  ``The universe is accelerating. Do we need a new mass scale?,''
  arXiv:0910.3949 [astro-ph.CO].


\bibitem{torsion}
  S.~Schlamminger, K.~Y.~Choi, T.~A.~Wagner, J.~H.~Gundlach and E.~G.~Adelberger,
  ``Test of the Equivalence Principle Using a Rotating Torsion Balance,''
  Phys.\ Rev.\ Lett.\  {\bf 100}, 041101 (2008)
  [arXiv:0712.0607 [gr-qc]].

  \bibitem{fedo1}
  F.~Piazza,
  ``Glimmers of a pre-geometric perspective,''
  arXiv:hep-th/0506124; F.~Piazza,
  ``Quantum degrees of freedom of a region of spacetime,''
  AIP Conf.\ Proc.\  {\bf 841}, 566 (2006)
  [arXiv:hep-th/0511285].

 \bibitem{fabio1}
 F.~Piazza and F.~Costa,
  ``Volumes of Space as Subsystems,''
  PoS(QG-Ph)032,  Proceedings of  ``From Quantum to Emergent Gravity: Theory and Phenomenology" [arXiv:0711.3048].

 \bibitem{sergio}
  S.~Cacciatori, F.~Costa and F.~Piazza,
  ``Renormalized Thermal Entropy in Field Theory,''
  Phys.\ Rev.\  D {\bf 79}, 025006 (2009)
  [arXiv:0803.4087 [hep-th]].
 
   \bibitem{paolo}
  P. Zanardi, D. A. Lidar and S. Lloyd
  ``Quantum tensor product structures are observable-induced'',
  Phys. Rev. Lett. \textbf{92}, 060402 (2004)
  [arXiv:quant-ph/0308043];
  
   
  \bibitem{fulling}
  L.~Parker and S.~A.~Fulling,
  ``Adiabatic regularization of the energy momentum tensor of a quantized field
  in homogeneous spaces,''
  Phys.\ Rev.\  D {\bf 9}, 341 (1974).



\bibitem{sarkar}
  A.~Blanchard, M.~Douspis, M.~Rowan-Robinson and S.~Sarkar,
  ``An alternative to the cosmological 'concordance model',''
  Astron.\ Astrophys.\  {\bf 412}, 35 (2003)
  [arXiv:astro-ph/0304237].

\bibitem{justin}
  N.~Afshordi, G.~Geshnizjani and J.~Khoury,
  ``Observational Evidence for Cosmological-Scale Extra Dimensions,''
  arXiv:0812.2244 [astro-ph].

 \bibitem{mark}
  J.~Khoury and M.~Wyman,
  ``N-Body Simulations of DGP and Degravitation Theories,''
  arXiv:0903.1292 [astro-ph.CO].

\bibitem{COBE}
 G.~Hinshaw {\it et al.},
 ``2-Point Correlations in the COBE DMR 4-Year Anisotropy Maps,''
  arXiv:astro-ph/9601061;
    D.~N.~Spergel {\it et al.}  [WMAP Collaboration],
 ``First Year Wilkinson Microwave Anisotropy Probe (WMAP) Observations:
  Determination of Cosmological Parameters,''
  Astrophys.\ J.\ Suppl.\  {\bf 148}, 175 (2003);
   C.~Copi, D.~Huterer, D.~Schwarz and G.~Starkman,
 ``The Uncorrelated Universe: Statistical Anisotropy and the Vanishing Angular
 Correlation Function in WMAP Years 1-3,''
  Phys.\ Rev.\  D {\bf 75}, 023507 (2007).

\bibitem{seljak}
 S.~Ho, C.~Hirata, N.~Padmanabhan, U.~Seljak and N.~Bahcall,
  ``Correlation of CMB with large-scale structure: I. ISW Tomography and
  Cosmological Implications,''
  Phys.\ Rev.\  D {\bf 78}, 043519 (2008). 

\bibitem{kash}
  A.~Kashlinsky, F.~Atrio-Barandela, D.~Kocevski and H.~Ebeling,
  ``A measurement of large-scale peculiar velocities of clusters of galaxies:
  technical details,''
  arXiv:0809.3733 [astro-ph]; arXiv:0809.3734 [astro-ph];
  R.~Watkins, H.~A.~Feldman and M.~J.~Hudson,
  ``Consistently Large Cosmic Flows on Scales of 100 Mpc/h: a Challenge for the
  Standard LCDM Cosmology,''
  arXiv:0809.4041 [astro-ph].
  



\end{thebibliography}
\end{document}